# An investigation of factors affecting Fast-Interaction Converter-driven Stability in Microgrids

Georgia Saridaki, Alexandros G. Paspatis, Panos Kotsampopoulos, Nikos Hatziargyriou

School of Electrical and Computer Engineering, National Technical University of Athens
Athens, Greece

*Abstract*-- **Massive integration of power electronic devices with multiple control schemes in a wide frequency range pose new challenges regarding systems stability and reliability. Interactions between the fast control loops or between the fast control loops and passive elements of the grid, have been reported in literature and have led to introducing a new type of stability: the Fast-Interaction Converter-driven Stability (FICDS). In this paper, factors affecting the FICDS, such as tuning controller parameters, line parameters, number of interconnected inverters, are explored in four microgrid topologies, operating in grid connected and islanded mode. With the use of an impedance based model which tracks the poles of the CDS transfer functions of each system, their stability has been assessed. The obtained results have been verified via time-domain simulations. Simulations from the islanded microgrid of Gaidouromantra in Greece showcase the impact of the control parameters on the operation of the system and indicate the need for further investigation.**

## I. INTRODUCTION

STABILITY of power systems has traditionally been a crucial issue in the planning and operation of both large and smaller power systems, such as the ones found in non-interconnected islands. Stability problems have been classified in voltage, frequency, and rotor angle stability in the classical reference [1]. With the wide deployment of renewable energy sources (RES) in the power systems to meet environmental goals [2], the dynamics affecting the operation of the power system have been altered, primarily due to the fact that most of the RES units are interfaced to the grid through power electronic devices. Inverter-based resources (IBRs) exhibit a different response than traditional generators directly coupled to the grid, owing to their fast reaction to dynamic phenomena and their associated control schemes. These features have led to new types of stability problems, classified as converter-driven stability, in the revised classification of power system stability [3], which has been extended to cover faster, electromagnetic types of phenomena.

Converter-driven related instability phenomena have been reported in several cases, e.g. oscillations occurred in high frequencies, from 500 Hz to 2kHz, in large wind farms connected with VSC-HVDC link [4] were classified as Fast-Interaction Converter-driven Stability phenomena (FICDS) [3], which was previously referred as harmonic instability [5]. Interactions between weak grid and STATCOM has also led to sub-synchronous and super-synchronous oscillations at the frequencies of 2.5 Hz and 97.5 Hz, respectively [6], and multiple parallel grid-connected inverters have been reported to interact and compromise system's stability by introducing new resonance frequencies [7]. To analyze stability, different tools based on classical control systems theory have been utilized over years. Most of the times, the state-space model of the system is developed, and stability analysis is performed based on the root-locus analysis [8]. Stability analysis can be also concluded by modelling the power system in the frequency domain [9]. To allow for easier adaptation to different power system cases, as well as to more complex systems, impedance-based modelling has recently been proposed [5], where the conclusions are drawn by evaluating the stability at each point of connection, rather than formulating the detailed state space model of the complete system. For this purpose, small signal linearized models of each component are extracted by using average techniques to remove switching discontinuities.

Microgrids, according to IEEE 2030.7 standard, are controllable small-scale entities comprising distributed energy resources, such as photovoltaic panels, wind turbines, battery energy storage systems, small diesel generators, and groups of loads. They can operate in grid connected and/or island mode, assuming that they have the necessary mechanisms and resources to ensure power supply, at least for their critical loads [10]. Several control strategies have been proposed in literature in order to integrate intelligent functions to inverter controllers for achieving effective voltage and frequency control. In any case, the operation of microgrids is based on extensive use of inverter interfaced resources, which makes them ideal cases for exploring the converter-driven stability issues [11].

Few papers have investigated FICDS in microgrids, while

This work was financially supported by the European Union's Horizon 2020 Research and Innovation Program and the Department of Science and Technology (DST), India through the RE-EMPOWERED Project under Grant Agreement No 101018420 and DST/TMD/INDIA/EU/ILES/2020/50(c) respectively.
Paper accepted to the International Conference on Power Systems Transients (IPST2023) in Thessaloniki, Greece, June 12-15, 2023.

they are mostly focused on grid connected mode and do not address the transition to islanded operation. Authors in [12], [13], [14] evaluate stability for grid connected systems consisting of multiple grid-following inverters, using several approaches such as a modified Nyquist criterion based on the global admittance encirclements at (0,j0) and a reformed global admittance criterion in which the real and imaginary part of the global admittance are plotted in the frequency domain. In [5], converter-driven stability is assessed by applying the Nyquist theorem to the minor feedback loop of each inverter in an islanded microgrid with two grid-following and one grid-forming inverter.

In this paper, impedance-based modelling is utilized to investigate FICDS in four case studies of microgrids, in order to draw conclusions about the effects of the presence of a stiff grid in interconnected mode, the length of the distribution line interconnecting the microgrid to the grid, the interaction between the interconnected inverters and the control schemes utilized, i.e. grid-following or grid-forming inverters, including their key parameters. Moreover, it investigates the transition of a microgrid from grid-connected to island mode with regard to FICDS phenomena. Preliminary simulation results from the microgrid in Kythnos island are presented. The paper is organized as follows: In Section II, the modelling for the power electronic converters is explained and Section III shows the methodology for FICDS investigation. Section IV presents the case studies with insights regarding FICDS in microgrids. Section VI concludes the paper.

## II. INVERTER MODELLING

### A. GRID-FOLLOWING INVERTER

The block diagram of the grid-following inverter is given in Fig. 1, as proposed in [5]. The PR controller $G_{pr}$ is responsible for regulating the output current of the inverter to the reference value $I_{ref}$. Parameter $G_d$ is introduced in the model to represent the delay due to the inverter's modulation PWM technique and the telecommunication systems. An LCL filter is used for minimizing the harmonics injected to the grid. The slow dynamics of the PLL and the outer high level control loops have been neglected.

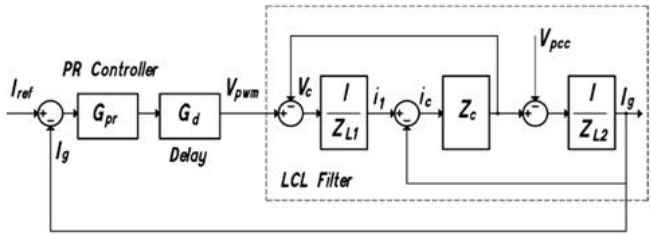

*Fig. 1. Block diagram of grid-following inverter*

$$G_{pr} = k_p + \frac{k_i s}{s^2+\omega_1^2}, \quad G_d = e^{-1.5T_s s} \quad (1)$$

The Norton equivalent of the inverter is illustrated in Fig. 2, whereas $Y_{oc}$, the output admittance of the inverter, is the transfer function between the first input signal $V_{pcc}$ and the output signal $I_g$. $G_c$ represents the dynamic performance of the grid-following inverter and is modelled as the transfer function between the second input signal $I_{ref}$ and the output signal $I_g$.

$$(I_{ref} - I_g) * G_{pr} * G_d = V_{pwm}, \quad (V_{pwm} - V_c) * \frac{1}{Z_{L1}} = i_1$$

$$i_1 - I_g = i_c = \frac{V_c}{Z_c}, \quad (V_c - V_{pcc})1/Z_{L2} = I_g \quad (2)$$

Using the system's equations (2) extracted by the block diagram of Fig. 1, $I_g$, $Y_{oc}$ and $G_c$ are calculated as follows:

$$I_g = I_{ref} * \left(\frac{G_d * G_{pr} * Z_C}{Z_{L1} * Z_c + Z_c * Z_{L2} + Z_{L2} * Z_{L1} + G_{pr} * G_d * Z_c}\right)$$

$$- V_{pcc} * \left(\frac{Z_{L1} + Z_c}{Z_{L1} * Z_c + Z_c * Z_{L2} + Z_{L2} * Z_{L1} + G_{pr} * G_d * Z_c}\right) \quad (3)$$

$$Y_{oc} = \frac{I_g}{V_{pcc}} = \frac{Z_c + Z_{L1}}{Z_{L1} * Z_c + Z_c * Z_{L2} + Z_{L2} * Z_{L1} + G_{pr} * G_d * Z_c} \quad (4)$$

$$G_c = \frac{I_g}{I_{ref}} = \frac{G_{pr} * G_d * Z_c}{Z_{L1} * Z_c + Z_c * Z_{L2} + Z_{L2} * Z_{L1} + G_{pr} * G_d * Z_c} \quad (5)$$

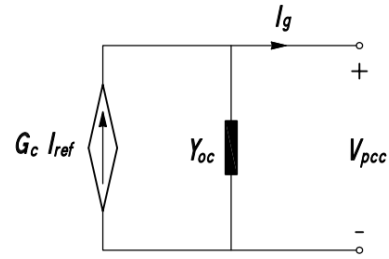

*Fig. 2. Norton equivalent of grid-following inverter*

### B. GRID-FORMING INVERTER

The block diagram of the grid-forming inverter is presented in Fig. 3, as proposed in [5]. The PR controller $G_{pr}$ is responsible for regulating the output voltage of the inverter to the reference value $V_{ref}$. The P controller $G_p$ is responsible for regulating the current of the inductance $i_L$ to the reference value $I_{ref}$, which is the output of the voltage control. The LC filter is connected to the output of the inverter for harmonic mitigation. The delay $G_d$ is introduced to the model due to the PWM technique and the telecommunication systems. The slow dynamics of the PLL and the outer high level control loops have been neglected.

$$G_{pr} = k_p + \frac{k_i s}{s^2+\omega_1^2}, \quad G_p = k_p, \quad G_d = e^{-1.5T_s s} \quad (6)$$

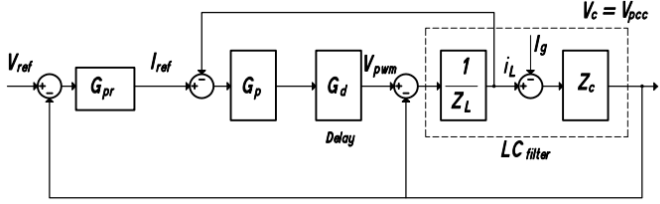

*Fig. 3. Block diagram of grid-forming inverter*

The Thevenin equivalent of the inverter is illustrated in Fig. 4, whereas $Z_{ov}$, the output impedance of the inverter, is the transfer function between the first input signal $I_g$ and the output signal $V_{pcc}$. $G_v$ models the dynamic performance of the grid-forming inverter which is the transfer function between the second input signal $I_{ref}$ and the output signal $V_{pcc}$.

$$(V_{ref} - V_c) * G_{pr} = I_{ref}, \quad (I_{ref} - i_L) * G_p * G_d = V_{pwm}$$

$$(V_{pwm} - V_c)\,^1/_{Z_L} = i_L, \quad i_L = \frac{V_C}{Z_C} + I_g \quad (7)$$

Using system's equations (7) extracted by the block diagram of Fig. 3, $V_{pcc}$, $Z_{ov}$ and $G_v$ are calculated as follows:

$$V_{pcc} = V_{ref} * \left(\frac{Z_C (G_d * G_p * G_{pr})}{Z_L + Z_c + G_{pr}*G_p*G_d + G_p*G_d}\right)$$

$$- I_g * \left(\frac{Z_C (G_d*G_p + Z_L)}{Z_L + Z_c + G_{pr}*G_p*G_d + G_p*G_d}\right) \quad (8)$$

$$Z_{ov} = \frac{V_{pcc}}{I_g} = \left(\frac{Z_C (G_d*G_p + Z_L)}{Z_L + Z_c + G_{pr}*G_p*G_d + G_p*G_d}\right) \quad (9)$$

$$G_v = \frac{V_{pcc}}{V_{ref}} = \left(\frac{Z_C (G_d*G_p*G_{pr})}{Z_L + Z_c + G_{pr}*G_p*G_d + G_p*G_d}\right) \quad (10)$$

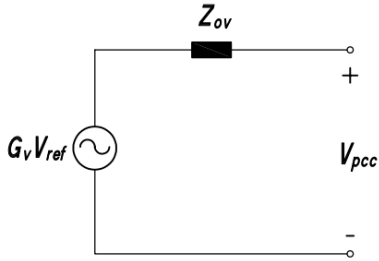

*Fig. 4. Thevenin equivalent of grid-forming inverter*

The above results for the equivalents of grid-following and grid-forming inverters, can be also obtained through classical control system analysis, by calculating the open loop transfer function of the system along with the transfer function between the output and each input when the loop of the other input is open [5].

## III. METHODOLOGY-ANALYSIS

Impedance based approach in the frequency domain is used in several publications which investigate FICDS. Several techniques have been introduced in literature to address the FICDS stability: Minor Loop Gain (MLG) and Global Minor Loop Gain (GMLG) aim to assess the stability of distribution grids by applying the Nyquist theorem to MLG and GMLG functions respectively [12] [15]. Global admittance (GA) aims to assess the stability by finding the zeros' location of the admittance sum of the system [12] [14]. In this paper, the stability is evaluated at the points of inverter connection by tracking the poles of CDS indices which are calculated as the transfer functions between the $V_{pcc}$ and the respective inputs of the power system under investigation.

### A. SINGLE INVERTER

1) *System A1: Grid connected – grid following inverter*

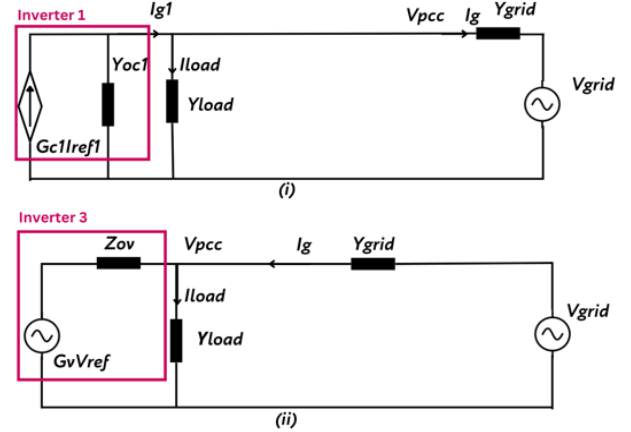

*Fig. 5. i) System A1: grid connected- grid-following ii) System A2: grid connected – grid-forming*

Grid-following inverters are connected to the grid and can provide ancillary services by integrating droops for voltage and frequency control [16]. To investigate their impact on the systems' FICDS stability, System A1 is considered and illustrated in Fig. 5i. Its respective block diagram is presented in Fig. 6i. The voltage at the point of connection is expressed in (12) in relation to the inputs $V_{grid}$, $I_{ref1}$ using system's state equations (11):

$$V_{pcc} = V_{grid} + Z_{grid} * I_g \quad, \quad I_g = G_{c1} * I_{ref1} - V_{pcc}(Y_{load} + Y_{oc1}) \quad (11)$$

$$V_{pcc} = \boldsymbol{V_{grid}} * \frac{1}{1 + Z_{grid} * Y_{oc1} + Z_{grid} * Y_{load}} + \boldsymbol{I_{ref1}}$$
$$* \frac{G_{c1} * Z_{grid}}{1 + Z_{grid} * Y_{oc1} + Z_{grid} * Y_{load}} \quad (12)$$

From (12), the following CDS indices are extracted:

$$CDS_1 = \frac{1}{1 + Z_{grid}*Y_{oc1} + Z_{grid}*Y_{load}},$$
$$CDS_2 = \frac{G_{c1}*Z_{grid}}{1 + Z_{grid}*Y_{oc1} + Z_{grid}*Y_{load}} \quad (13)$$

Applying the classical control theory at $CDS_1$ and $CDS_2$ (13), the FICDS stability can be evaluated by finding the location of the poles of CDS transfer functions. If at least one pole has a positive real part, then fast interaction converter driven instability is detected. Given that inverter 1 is stable and thus $G_{c1}$ transfer function does not have any pole in the positive real axis, the FICDS stability for system A1 can be evaluated by finding the location of the $CDS_1$ poles.

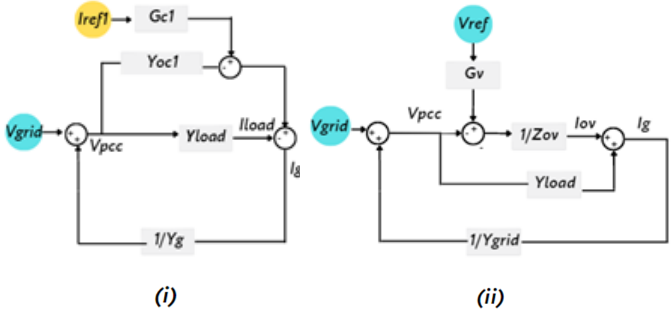

Fig. 6. Block diagrams: i) System A1 ii) System A2

A model with $CDS_1$ indice was developed in MATLAB for system A1 with the nominal parameters of Table I. Firstly, a parametric analysis with different values for the proportional gain of the PR current controller ($K_{p1}$) of inverter 1 was conducted by plotting the pole zero map for the $CDS_1$ index. It can be seen in Fig. 7i that when $K_{p1}$ increases from 7 to 7.5, System A1 becomes unstable. Moreover, the interaction between the PR controller with the grid is explored by tracking the location of the poles for the $CDS_1$ indices for different values of Ls when $K_{p1}$ is set to 7. As illustrated in Fig. 7ii, with the increase of the inductance of the grid from 0.3mH to 0.5mH poles with positive real part appear, rising instability issues.

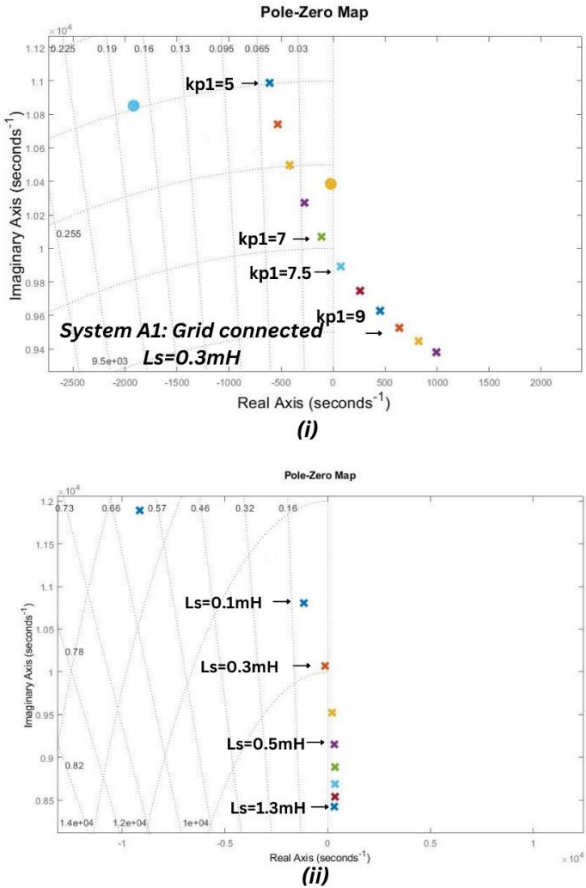

Fig. 7. CDS1 pole zero map for : i) different $k_{p1}$ values and ii) different grid conditions

*2) System A2: Grid connected – grid forming inverter*

When the mode of the inverter changes from grid-following to grid-forming (System A2 Fig. 5ii) a new block diagram of the system is designed in Fig. 6ii. The voltage at the point of connection is calculated in relation to $V_{grid}$, $V_{ref}$ (14) and the indicators are modified accordingly (15):

$$V_{pcc} = - V_{grid} * \frac{1}{1 + Z_{grid}*Y_{ov} + Z_{grid}*Y_{load}} + V_{ref}$$
$$* \frac{G_v}{1 + Z_{grid}*Y_{ov} + Z_{grid}*Y_{load}} \quad (14)$$

$$CDS_3 = \frac{1}{1+Z_{grid}*Y_{ov}+Z_{grid}*Y_{load}}, \quad CDS_4 = \frac{G_v}{1+Z_{grid}*Y_{ov}+Z_{grid}*Y_{load}} \quad (15)$$

Given that inverter 3 is stable and thus $G_v$ transfer function does not have any pole in the positive real axis, the CDS stability for system A2 can be evaluated by finding the location of $CDS_3$ indices in MATLAB. It is observed in Fig. 8, that if the inverter operates in grid-following mode (inverter 1) and Ls is 5mH, $CDS_1$ has poles with positive real part. When the mode of inverter changes to grid-forming (inverter 3), all the poles of $CDS_3$ move into the negative real half plane and thus the system becomes stable.

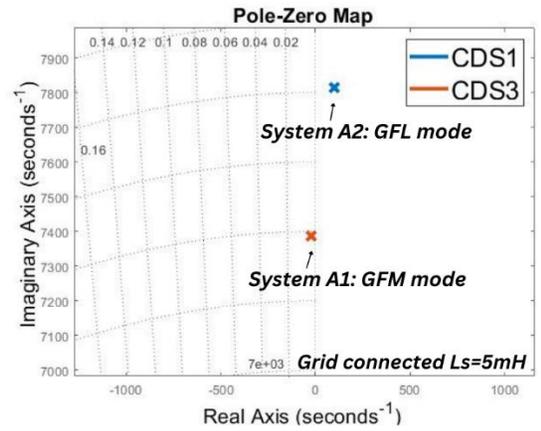

Fig. 8. CDS1 and CDS3 pole zero map for Ls=5mH

### B. PARALLEL INVERTERS

*1) System B1: 2 grid-following inverters- grid connected mode*

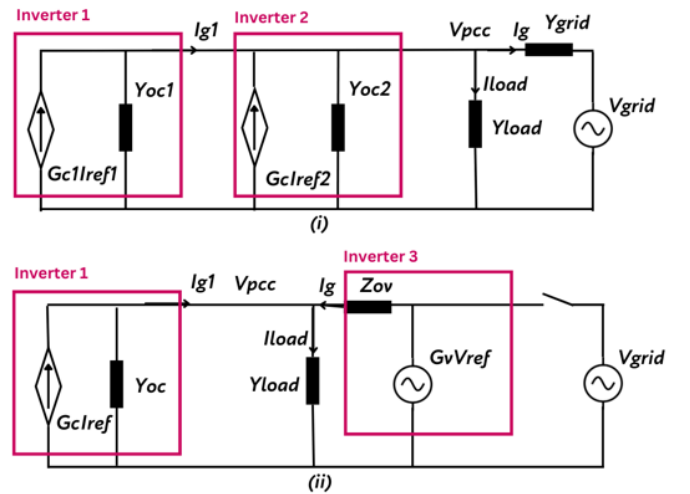

Fig. 9. i) System B1: 2 grid-following inverters- grid connected mode ii)

*System B2: grid-following inverter in parallel with grid-forming inverter: islanded operation*

To investigate the interactions between the inverters, the FICDS was firstly assessed in System B1 (Fig. 9i), a microgrid operating in grid connected mode consisting of two parallel inverters in grid-following mode (Inverter 1 and Inverter 2) with the parameters of Table I. The block diagram of System B1 is presented in Fig. 10i.

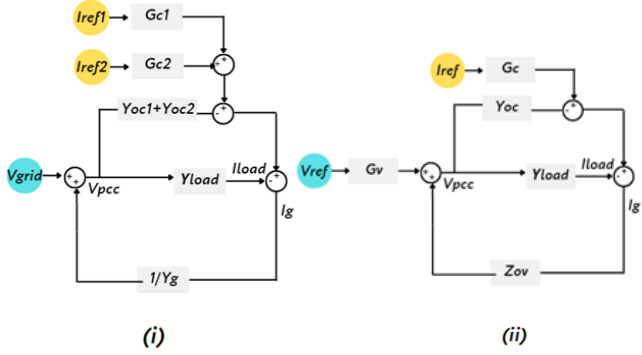

Fig. 10. Block diagram: i) System B1 ii) System B2

The voltage at the point of connection is expressed in relation to $V_{grid}$, $I_{ref1}$, $I_{ref2}$ as:

$$V_{pcc} = V_{grid} * \frac{1}{1 + Z_{grid} * Y_{oc1} + Z_{grid} * Y_{oc2} + Z_{grid} * Y_{load}} + I_{ref1}$$
$$* \frac{G_{c1} * Z_{grid}}{1 + Z_{grid} * Y_{oc1} + Z_{grid} * Y_{oc2} + Z_{grid} * Y_{load}} + I_{ref2}$$
$$* \frac{G_{c2} * Z_{grid}}{1 + Z_{grid} * Y_{oc1} + Z_{grid} * Y_{oc2} + Z_{grid} * Y_{load}} \quad (16)$$

The number of CDS indices are equal to the number of the system's inputs and are calculated as:

$$CDS_5 = \frac{1}{1+Z_{grid}*Y_{oc1}+Z_{grid}*Y_{oc2}+Z_{grid}*Y_{load}},$$
$$CDS_6 = \frac{G_{c1}*Z_{grid}}{1+Z_{grid}*Y_{oc1}+Z_{grid}*Y_{oc2}+Z_{grid}*Y_{load}},$$
$$CDS_7 = \frac{G_{c2}*Z_{grid}}{1+Z_{grid}*Y_{oc1}+Z_{grid}*Y_{oc2}+Z_{grid}*Y_{load}} \quad (17)$$

Given that both inverters are stable and thus $G_{c1}$, $G_{c2}$ transfer functions do not have any pole in the positive real axis, the FICDS stability for system B1 can be evaluated by finding the location of $CDS_5$ indices. MATLAB is used for the extraction of the poles for the $CDS_5$ transfer function. By plotting the $CDS_5$ pole zero map for different values of the proportional gain of the PR current controller of inverter 1 ($k_{p1}$), illustrated in Fig. 11i, it is revealed that system B1 becomes unstable when $k_{p1}$ changes from 6 to 6.5. It is observed that for the same inverter, the range of $k_{p1}$ for stable operation changes and is smaller than the respective one for system A1, revealing the impact of the number of the interconnecting inverters on their tuning parameters.

### 2) System B2: grid-following inverter in parallel with grid-forming inverter: islanded operation

In case of unexpected disturbances or faults resulting in the activation of protective devices and disconnection of the main grid, a microgrid can operate islanded to meet the load demand (System B2). In this case which is illustrated in Fig. 9ii at least one of the connected inverters must transition from grid-following to grid-forming mode and will be responsible for keeping the voltage and the frequency of the microgrid within the permitted limits. The revised block diagram of System B2 is given in Fig. 10ii. The voltage at the point of connection in relation to $V_{ref}$, $I_{ref}$, and the CDS indices are presented in (18) and (19).

$$V_{pcc} = V_{ref} * \frac{G_v}{1 + Z_{ov} * Y_{oc1} + Z_{ov} * Y_{load}} + I_{ref}$$
$$* \frac{G_{c1} * Z_{ov}}{1 + Z_{ov} * Y_{oc1} + Z_{ov} * Y_{load}} \quad (18)$$

$$CDS_8 = \frac{G_v}{1+Z_{ov}*Y_{oc1}+Z_{ov2}*Y_{load}}, \quad CDS_9 = \frac{G_{c1}*Z_{ov}}{1+Z_{ov}*Y_{oc1}+Z_{ov}*Y_{load}} \quad (19)$$

Given that $G_{c1}$ transfer function does not have any pole in the positive real axis, the FICDS stability for system B2 can be evaluated by finding the location of $CDS_8$ indices.

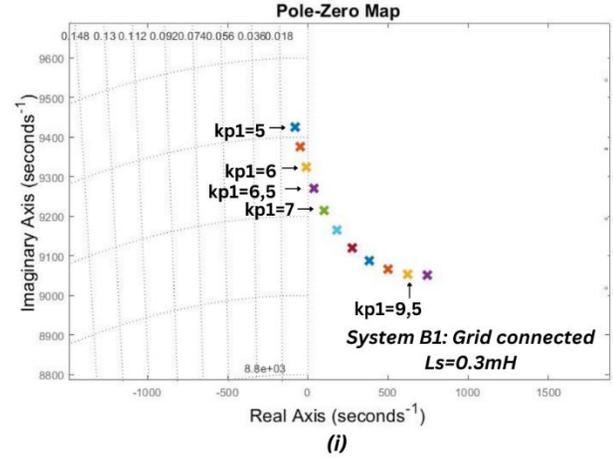

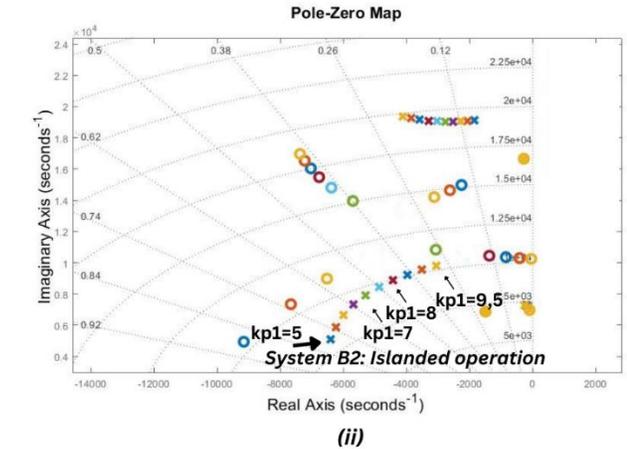

Fig. 11. i) CDS5 pole zero map for different kp1 values ii) CDS8 pole zero map islanded operation

For system B2 with the parameters of Table I, it can be seen in Fig. 11ii that $CDS_8$ has no poles in the positive right half plane for a range of $k_{p1}$ starting from 5 to 9.5 which reveals that the grid-following inverter can now operate with bigger proportional gain that in the previous test cases. Nevertheless, when the proportional gain of the PR controller for the voltage control of the grid-forming inverter $k_{p3}$ increases from 0.1 to 0.2, the $CDS_8$ indices demonstrate multiple poles in the right half plane and the thus system is unstable.

TABLE I
SYSTEMS INITIAL PARAMETERS

| Systems initial parameters | Values |
|---|---|
| *Inverter 1: Grid-following mode* | |
| **Current controller (PR)** | |
| $K_{p1}$ | 7 |
| $K_{i1}$ | 1000 |
| **Filter LCL** | |
| L | 1.2mH/0.1 Ω |
| C | 15μF |
| L | 0.3mH/0.2Ω |
| *Inverter 2: Grid-following mode* | |
| **Current controller (PR)** | |
| $K_{p2}$ | 5 |
| $K_{i2}$ | 1000 |
| **Filter LCL** | |
| L | 1.5mH/0.1 Ω |
| C | 15μF |
| L | 0.5mH/0.2Ω |
| *Inverter 3: Grid-forming mode* | |
| **Voltage controller (PR)** | |
| $K_{p3}$ | 0.1 |
| $K_{i3}$ | 100 |
| **Current Controller (P)** | |
| $K_{p4}$ | 5 |
| **Filter LC** | |
| L | 1.5mH/0.1 Ω |
| C | 28μF |
| *Grid* | |
| $R_S$ | 0.4 Ω |
| $L_S$ | 0.3mH |
| $V_{grid\ (rms)}$ | 230V |
| Load | 100 Ω |
| Sampling frequency | 10kHz |
| System frequency | 50Hz |

## IV. SIMULATION RESULTS

### A. SINGLE INVERTER

For validation of the results obtained from MATLAB based on the theoretical analysis, time-domain simulation models for systems A1 and A2 were built in the SIMULINK environment. Firstly, it can be seen in Fig. 12i, that for system A1 with the nominal parameters of Table I, the voltage at the common point of connection is stable. At $t_1= 5s$, the $kp_1$ is increased from 7 to 7,5 and instability issues occur (Fig. 12i). Then, $kp_1$ is restored to 7 and $L_s$ changes from 0.3mH to 5mH at $t_2=8s$. It can be observed in Fig. 12ii, that in this case, instability arises at the voltage at the point of connection. After modifying the control scheme of the inverter from grid-following to grid-forming, at $t_3=8s$, the results of the MATLAB model have been verified and the system becomes stable.

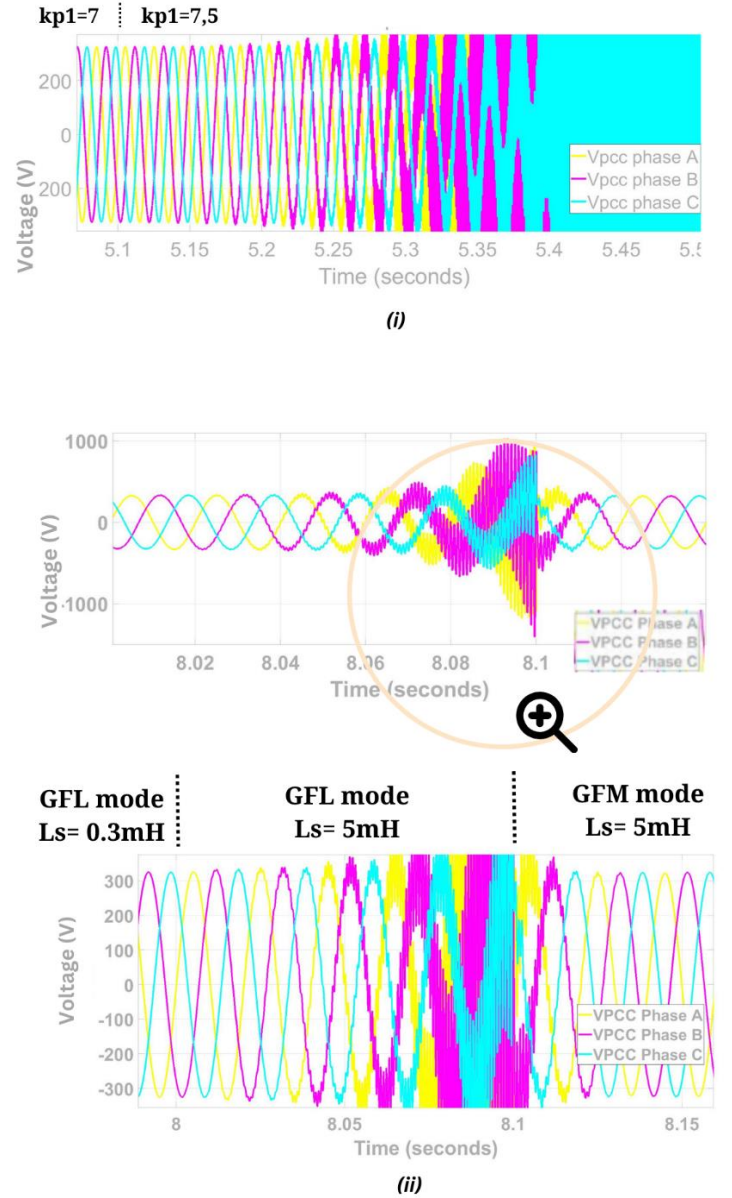

*Fig. 12. Voltage at the point of connection $V_{pcc}$ i) kp1 increase ii) transition to grid-forming*

The performed scenarios reveal that weak interconnection with the main grid leads to fast-interactions converter driven instability issues in System A1. Hence, if a new inverter is planned to be installed, the location of the point of connection should be thoroughly investigated. Additionally if the current controller of the grid-following inverter is not tuned correctly and exceeds the value of 7, System A1 becomes unstable. When the inverter operates in grid-forming mode (System A2), the stable operation in weak interconnection conditions is restored, proving the applicability of grid-forming control schemes.

## B. PARRALEL INVERTERS

A simulation was developed in the SIMULINK environment for Case B1 and B2 which verifies the results of the theoretical analysis. In Fig. 13i the voltage at the common point of connection is presented. Firstly, when $k_{p1}=6$ and $k_{p2}=5$ the system is stable and at $t_1=3s$ when the proportional gain of Inverter 1 increases from 6 to 7, system demonstrates unstable behavior (Fig. 13ii).

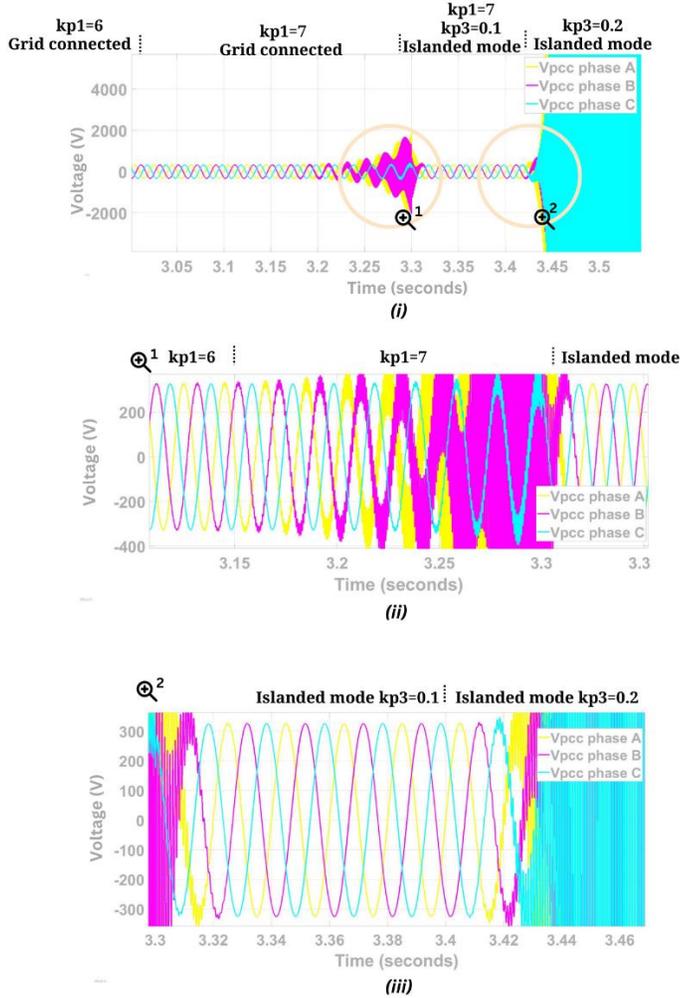

*Fig. 13. i) Voltage at the point of connection ii) kp1 increase and transition to inslanded iii) kp3 increase*

At $t_2=3.3s$ the connection with the grid is lost and the microgrid is islanded; one of the inverters operates in grid-forming mode (Inverter 3) while the other remains in grid-following mode (Inverter 1). It can be seen in Fig. 13ii that the system exhibits now a stable behavior. When the proportional gain of the PR voltage controller of the grid-forming inverter changes from 0.1 to 0.2 at $t_3=3.4s$ (Fig. 13iii), severe instability phenomena occurs. Again, the results match well with the theoretical analysis of the previous section.

From these scenarios, it is seen that interactions between the current control of the grid-following inverters lead to instability issues in System B1. Reduced proportional gains $k_{p1}$ of the current control of the grid-following inverters restore the system to the stable operation. The number of the grid-connected inverters plays a key role to the stability of the system. Transition from grid connected to islanded (System B2) results in the stable operation of the microgrid but if the PR controller for the voltage control of the grid-forming inverter in not tuned correctly, stability problems will arise again.

## C. GAIDOUROMANTRA MICROGRID

Gaidouromantra is the first microgrid in Europe, installed in 2001 to electrify 14 vacation houses in Kythnos island, Greece [17]. The microgrid is permanently islanded and it is supplied by distributed energy resources, such as photovoltaic panels, a wind turbine and a battery energy storage system. Six photovoltaic systems are installed producing a total of 11.145 kWp. Each PV installation is equipped with grid-following inverters as illustrated in Fig. 14. Voltage and frequency are controlled by the 3 single phase battery grid-forming inverters of the battery bank with nominal capacity 11900Ah/48V that is located in the System house and performs intelligent load control. The grid-forming inverters also employ droop control for power management. A back up diesel generation of 22kVA is also available. It is currently being upgraded in the framework of RE-EMPOWERED project, a collaboration between two continents, Europe and India [18]. This joint research project aims to provide solutions for multi-energy islands and microgrids, which will be implemented in four demo sites in Bornholm (Europe), Gaidouromantra (Europe), Ghoramara (India) and Keonjar (India).

A model in SIMULINK was developed for emulating the operation of Gaidouromantra microgrid. It can be seen in Fig. 15, that when the proportional gain of one PID current controller of a grid-following inverter increases from 20 to 201, the output current of the grid-following inverter along with the voltage at the System house become unstable and the system collapses. These initial results reveal the impact of tuning the parameters of the inverters control on the stability of the system. More detailed studies including the extraction of the corresponding CDS indices for the microgrid taking into consideration the length of the lines and the different point of connections, coupled with Hardware-in-the-Loop simulations and on-site measurements are in progress.

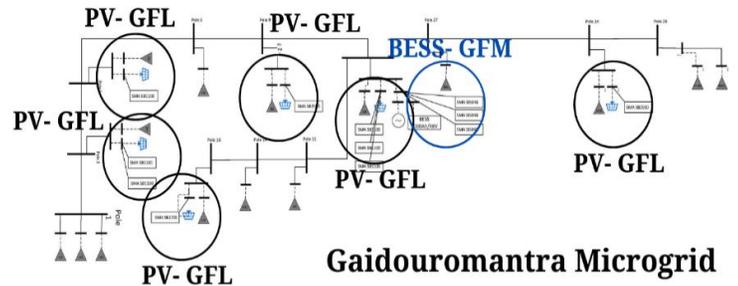

*Fig. 14. Gaidouromantra microgrid*

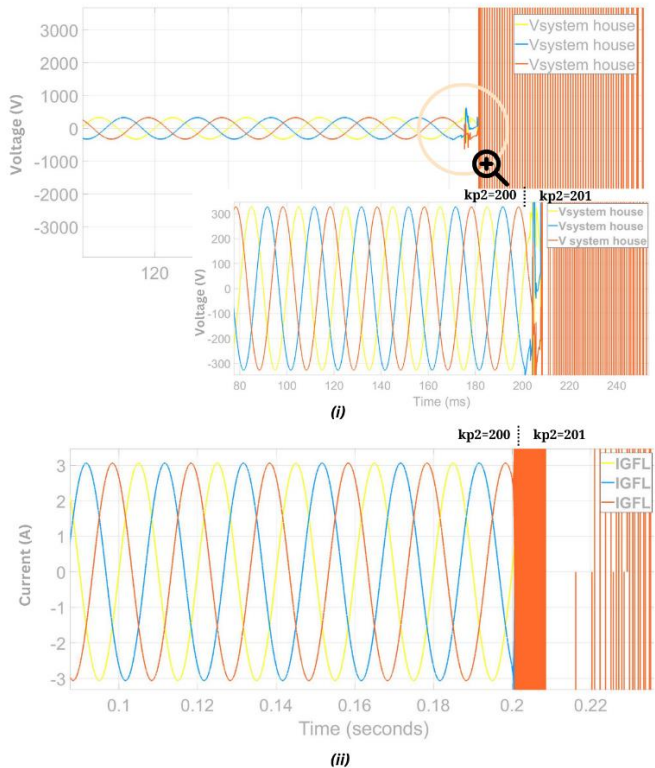

*Fig. 15. i) Voltage at the System House ii) Output current of the grid-following inverter*

## V. CONLCUSIONS

This paper investigates factors affecting the Fast-Interaction Converter-driven Stability in four simple microgrid topologies with the use of the impedance based analysis. The CDS set of indices extracted for each system is used to predict when a fast-interaction converter-driven instability occurs. The results of the MATLAB model are verified via simulations in SIMULINK software.

Overall, it is observed that the increase of the length and consequently the impedance of the line which connects the microgrid with the main grid, can raise instability problems which are resolved if the inverter changes its mode of operation from grid-following to grid-forming. The number of interconnected inverters also plays a key role to this type of stability and the same inverter has different limits for its tuning parameters depending on the microgrid topology. Lastly, the transition of a microgrid to islanded mode can enable grid-following inverters to operate with larger tuning parameters but can also potentially lead to instability, if the grid-forming is tuned incorrectly.

ACKNOWLEDGMENT

This work was financially supported by the European Union's Horizon 2020 Research and Innovation Program and the Department of Science and Technology (DST), India through the RE-EMPOWERED Project under Grant Agreement No 101018420 and DST/TMD/INDIA/EU/ILES/2020/50(c) respectively.